\documentclass[preprint]{elsarticle}

\usepackage{textcomp}
\usepackage{verbatim}
\graphicspath{figs}
\usepackage{algpseudocode}
\usepackage{epstopdf}
\usepackage{color}
\usepackage{listings}
\usepackage{hyperref}
\lstdefinestyle{customfortran}{
  language=[90]Fortran,
  basicstyle=\footnotesize\ttfamily,
  showstringspaces=false,
  morecomment=[l]{!\ }
}
\lstdefinestyle{customxml}{
  belowcaptionskip=1\baselineskip,
  xleftmargin=\parindent,
  language=XML,
  basicstyle=\footnotesize\ttfamily,
  showstringspaces=false,
}
\lstdefinestyle{customrnc}{
  belowcaptionskip=1\baselineskip,
  xleftmargin=\parindent,
  language=Python,
  basicstyle=\footnotesize\ttfamily,
  showstringspaces=false
}

\def\tag#1{\textsf{$<$#1$>$}}
\def\att#1{\textsf{#1}}
\def\arg#1{\textsl{#1}}
\def\name#1{\textsf{#1}}
\def\routine#1{\textsf{#1}}

\newcommand{\libpsml}[1]{\textsl{libPSML}#1}

\journal{Computer Physics Communications}

\begin{document}

\begin{frontmatter}

\title {The {\sc psml} format and library for norm-conserving 
        pseudopotential data curation and interoperability}

\address[icmab]{ Institut de Ci\`encia de Materials de Barcelona (ICMAB-CSIC),
              Campus UAB, 08193 Bellaterra, Spain}

\author[icmab]{Alberto Garc\'{\i}a\corref{cor1}}
\ead{albertog@icmab.es}
\cortext[cor1]{Corresponding author}
\author[liege]{ Matthieu Verstraete }
\ead{Matthieu.Verstraete@ulg.ac.be}
\address[liege]{nanomat/Q-MAT/CESAM,
              Universit\'e de Li\`ege, All\'ee du 6 Ao\^ut 19 (B5a),
              B-4000 Li\`ege, Belgium}
\author[unican]{ Yann Pouillon }
\ead{yann.pouillon@unican.es}
\author[unican]{ Javier Junquera }
\ead{javier.junquera@unican.es}
\address[unican]{ Departamento de Ciencias de la Tierra y
              F\'{\i}sica de la Materia Condensada, Universidad de Cantabria,
              Cantabria Campus Internacional,
              Avenida de los Castros s/n, 39005 Santander, Spain}
\date{\today}

\begin{abstract}
Norm-conserving pseudopotentials are used by a significant number of
electronic-structure packages, but the practical differences among
codes in the handling of the associated data hinder their
interoperability and make it difficult to compare their results. At
the same time, existing formats lack provenance data, which makes it
difficult to track and document computational workflows.
To address these problems, we first propose a
file format ({\sc psml}) that maps the basic concepts of the norm-conserving
pseudopotential domain in a flexible form and supports the inclusion
of provenance information and other important metadata. Second, we
provide a software library (\libpsml{}) that can be used by electronic structure
codes to transparently extract the information in the file and adapt
it to their own data structures, or to create converters for other
formats. Support for the new file format has
been already implemented in several pseudopotential generator programs
(including {\sc atom} and {\sc oncvpsp}), and the library has been linked with
{\sc siesta} and {\sc abinit}, allowing them to work with 
the same pseudopotential
operator (with the same local part and fully non-local projectors) 
thus easing the comparison of their results for the structural and
electronic properties, as shown for several example systems. This methodology can
be easily transferred to any other package that uses norm-conserving
pseudopotentials, and offers a proof-of-concept for a general approach
to interoperability.

\begin{flushleft}
\end{flushleft}

\end{abstract}

\end{frontmatter}

{\bf PROGRAM SUMMARY}

\begin{small}
\noindent
{\em Program Title:} \libpsml{}                               \\
{\em Licensing provisions:} BSD 3-clause                           \\
{\em Programming language:} Fortran     \\
{\em Distribution format:} tar.gz                              \\
{\em External routines/libraries:} xmlf90 for XML handling in Fortran
(http://launchpad.net/xmlf90) \\

{\em Nature of problem:} \\
Enhancing the interoperability of electronic-structure codes by
sharing pseudopotential data.
   \\
{\em Solution method:}\\
  Create an XML-based pseudopotential format ({\sc psml}), complete with a
  formal schema, and a processing library (\libpsml{}) that transparently
  connects client codes to the information in the format.
   \\

\end{small}

\section{Introduction}

Within computational science, reproducibility of research goes beyond
using a specific version of a code and the appropriate input
files. What is really sought is to replicate a certain physical result
with a different code which implements the same basic equations of the
domain at hand, but with a different set of approximations or details
of implementation. This latter code will most likely have a different
input data format, which might not be perfectly mapped to the format
used by the original code. Reproducibility is still possible if
the input data are curated so that their essential ontological
properties are preserved, since it can be assumed that properly
implemented codes within a domain will share a basic ontology for this
domain, and can properly interpret its elements.

Here we are concerned with electronic structure methods in the broad
sense~\cite{Martin,Kohanoff}, and in particular density functional
theory (DFT~\cite{Hohenberg-64,Kohn-65}), which have become standard
theoretical tools to analyze and explain experiments in chemistry,
spectroscopy, solid state physics, material science, biology, and
geology, among many other fields.

Programs which implement DFT fall into two main categories, those
which treat all electrons explicitly (linear augmented plane wave,
LAPW~\cite{Andersen-75,Singh-lapw}, linear muffin tin orbitals,
LMTO~\cite{Andersen-75,Skriver-lmto}, or the Korringa-Kohn-Rostoker
and related multiple-scattering theory methods,
KKR~\cite{Korringa-47,Kohn-54}), and those which replace the effect of
the chemically inert core electrons by a (usually non-local)
pseudopotential~\cite{Phillips-59, Harris-78, Hamann-79}.

After some historical initial attempts, seminal work by Hamann,
Schl{\"u}ter and Chiang~\cite{Hamann-79} and Kerker~\cite{Kerker-80}
showed how \emph{norm-conserving pseudopotentials} achieve a good
tradeoff between \emph{transferability} (a pseudopotential constructed
for a specific environment, usually the isolated atom, gives good
results when the atom is placed in a different environment) and
\emph{softness} (as measured for example by the number of plane waves
that must be used in the representation of the wave functions for a
good convergence of the physical properties).
Since the 1990s more sophisticated schemes have been developed to
treat the basic problem of eliding the core electrons.  In particular,
ultrasoft pseudopotentials~\cite{Vanderbilt-90}, and the Projector
Augmented Wave (PAW) method~\cite{Bloch-94} are in widespread use 
for their improved accuracy and features, although they involve a
significantly higher degree of complexity in the implementation and
maintenance of the algorithms for electronic structure determination
and analysis.

Several well-known atomic DFT
programs~\cite{FHI-PP,APE,Opium,Froyen,oncvpsp,QE-atomic} generate
pseudopotentials in a variety of formats, tailored to the needs of
electronic-structure codes. While some generators are now able to
output data in different bespoke formats, and some simulation codes
are now able to read different pseudopotential formats, the common
historical pattern in the design of those formats has been that a
generator produced data for a single particular simulation code, most
likely maintained by the same group. This implied that a number of
implicit assumptions, shared by generator and user, have gone into the
formats and fossilized there.  Examples include, among others, many
flavours of radial grids (from linear~\cite{oncvpsp,Hamann-13} to
logarithmic~\cite{Froyen,Troullier-91}, including all kinds of
powers~\cite{Teter-93} or geometrical series~\cite{FHI-PP,Fuchs-99}),
different ways of storing radial functions with spherical symmetry
(angularly integrated with $4 \pi r^{2}$ factors included, where $r$
stands for the distance to the nuclei, or multiplication by different
powers of $r$ considered), different normalization conditions, etc.
This leads to practical problems, not only of programming, but of
interoperability and reproducibility, which depend on spelling out
quite a number of details which are not well represented for all codes
in existing formats.

Moreover, pseudopotential information can be produced and used at
various levels. The original work was based on a set of semi-local
operators (non-local in the angular part and local in the radial
part), whereas very soon a computationally more efficient form based
on fully non-local projectors plus a local part~\cite{Kleinman-82} was
developed, and is now the standard norm-conserving form used by
electronic-structure codes.  The transformation of a semi-local
pseudopotential into a fully non-local operator is not univocally
defined from the semi-local components alone.  Therefore, different
codes can yield different results even if reading the same semi-local
information from an input file. For example, most plane waves codes
make the so-called local part equal to one of the semilocal components
for reasons of efficiency, whereas {\sc siesta} optimizes the local
part for smoothness, since it is the only pseudopotential part that
needs to be represented in a real space grid~\cite{Soler-02}. Modern
pseudopotential generators are able to produce directly a set of
non-local operators plus a local part.  True interoperability can be
achieved only if the codes use the same final projector-based
pseudopotential.

Straightfoward interoperability of codes would allow all to benefit from the best
capabilities of each.  On the one hand, removing the
variability associated with the pseudopotential would enable a direct
test of the quality of the basis sets used in the expansion of the
one-particle Kohn-Sham eigenstates by different codes (with
identical pseudopotentials, and at convergency, one should get the
same total energy for the same atomic structure with any code).
On the other hand, we can imagine a situation where the most stable
atomic configuration of a large system can be found with a given code
in a cheap yet accurate way, and then passed to another for further
analysis through single-shot expensive calculations for the fixed
geometry.

The obvious benefits of interoperability have naturally spawned
efforts at providing more appropriate data mechanisms. At the most
basic level, some pseudopotential generators (e.g. {\sc oncvpsp} or {\sc ape}) offer options
to write different output files suitable for different
electronic-structure codes. While this addresses some of the problems,
it falls short of a fully satisfactory solution. More robust are
efforts to provide a standard format that can be used
universally. Within the PAW domain, the PAW-XML format~\cite{paw-xml}
comes close to this goal, being produced by a number of
PAW dataset generators and read by most PAW-enabled
electronic-structure codes. The UPF (Unified Pseudopotential
Format)~\cite{upf} is meant to encompass the full range of
pseudopotential options, including (semi)local and fully-non-local
norm-conserving, ultrasoft pseudopotentials, and PAW datasets. It
is used within the Quantum Espresso suite of codes and converters
exist for other codes.

We believe that, indeed, the solution to the interoperability problems
involves the design a data format that faithfully maps the relevant
concepts of the domain's ontology at all the relevant levels
(semi-local pseudopotentials, charge densities, non-local projectors,
local potentials, etc). But the format must also provide appropriate
metadata that represents provenance (generation and any further
processing) and documents in a parseable form any details that might
be needed downstream. This second aspect has not been properly addressed
so far.

The need for standardisation of the pseudopotential format and the
provision of richer metadata to track provenance and document
computational workflows has been made more relevant by the appearance
during the last few years of high-throughput simulation schemes for
materials design~\cite{materials-project,Ceder-13, AiiDA}, which need
well-tested and efficient pseudopotential libraries to draw upon.
Examples of the latter are the ultrasoft pseudopotential library by
Garrity, Bennett, Rabe and Vanderbilt~\cite{Garrity-14,GBRV}, the
PSLibrary associated with QE~\cite{PSlibrary}, the
Jollet-Torrent-Holzwarth~\cite{jth} PAW library, and the libraries
being built~\cite{Schlipf-15,gygi,pseudo-dojo} using multi-projector
norm-conserving pseudopotentials generated by the {\sc oncvpsp}
code~\cite{Hamann-13}. The latter are proving competitive with
ultrasoft pseudopotentials and the PAW method in
accuracy.~\cite{gygi,pseudo-dojo}

Motivated by the all the previous considerations, in this paper we
present a file format for pseudopotential data ({\sc psml}, for
PSeudopotential Markup Language) which is designed to encapsulate as
much as possible the abstract concepts in the domain's ontology, and
to provide appropriate metadata and provenance information.  Moreover,
we provide a software library (\libpsml{}) that can be used by electronic
structure codes to transparently extract the information in a {\sc psml}
file and adapt it to their own data structures, or to create
converters for other formats.  Our initial focus is the sub-domain in
which norm-conserving pseudopotentials are used, which is not
restricted to legacy cases but is set to grow in importance due to the
new multi-projector pseudopotentials mentioned above.
As the format is based on XML (eXtensible Markup
Language)~\cite{xml}, it is very flexible and can serve as a basis for
the future accomodation of PAW datasets and ultrasoft
pseudopotentials.  Our work falls within the scope of
CECAM's Electronic Structure Library (ESL) initiative~\cite{ESL},
which aims at building a collection of software functionalities,
including standards and interfaces, to facilitate the development of
electronic structure codes.

This paper is organized as follows: Section~\ref{sec:psml-format}
describes the basic elements and structure of a {\sc psml} file, with a more
formal XML schema given in Sec.~\ref{sec:schema}. The \libpsml{} library
is described in Sec.~\ref{sec:libpsml}. Section~\ref{sec:discussion}
discusses the relationship of the current work to
previous efforts, and provides a guide to the tools already available in the
{\sc psml} ecosystem and the mechanisms available for development of new
ones. Finally, as an example of the interoperability benefits afforded
by {\sc psml}, we present in Sec.~\ref{sec:results} a comparison of actual
physical results obtained by {\sc abinit} and {\sc siesta} using the same
{\sc psml} files.

\section{The {\sc psml} file format}
\label{sec:psml-format}

This section contains a design rationale and human-readable
description of the format. A formal schema specification can be found
in Sec.~\ref{sec:schema}. The following documents version 1.1 of the
{\sc psml}, current as of this writing. Any updates and complementary
information are available at the ESL {\sc psml} site:
\textsl{http://esl.cecam.org/PSML}.

\subsection{Root element}
The root element is \tag{psml}, containing a \att{version} attribute
for use by parsers, and two attributes to make explicit the
(mandatory for now) units used throughout the file: \att{energy-unit}
(hartree) and \att{length-unit} (bohr). In addition, the root element
should contain a \att{uuid} attribute to hold a {\it universally unique
identifier}~\cite{uuid}.

\subsection{Provenance element(s)}

The file should contain metadata concerning its own origin, to
aid reproducibility. As a minimum, it should contain information
about the program(s) used to generate or transform the
pseudopotential, 
ideally with version numbers and compilation options,
and provide a copy of any input files fed into the program(s). The
information is contained in \tag{provenance} elements (an 
example is provided in Table~\ref{table:provenance}). Its internal
structure is subject to a minimal specification:

\begin{itemize}
\item The attribute \att{creator}, where the name of the generator code 
is specified, is mandatory.
\item The attribute \att{date} is mandatory.
\item If input files are provided, they should be included as
  \tag{input-file} elements with the single attribute \att{name} and
  no children other than character data, which should be placed in a
  CDATA section to avoid processing of XML reserved characters.  For
  obvious reasons, the name of the file and its detailed content will
  depend on the program used to generate the pseudopotential. The name
  should be a mnemonic reference.
\item The inclusion of an \tag{annotation} element, providing arbitrary
  extra information in the form of key-value pairs, is encouraged. See
  the description and motivation of the \tag{annotation} element in
  Sec.~\ref{sec:annotations}.
\end{itemize}

There can be an arbitrary number of \tag{provenance} elements, ordered
in temporal sequence in the file, with the most recent first. Since
some XML processors might not preserve the order of the
elements, it is suggested that a \att{record-number} attribute be added to
make the temporal ordering explicit, with ``1'' for the oldest
operation, ``2'' for the second oldest, and so on.

This feature supports the documentation of successive
actions taken on the file's information. For example, a
pseudopotential-generation program might generate a {\sc psml} file
containing only semilocal potentials. This file is then processed by
another program that generates a local potential and the corresponding
non-local projectors. The {\sc psml} file produced by the second
program should keep the original provenance element, and add another
one detailing the extra operations taken.

 \begin{table}[h]
    \caption[ ]{ An example of the \tag{provenance} element }

\begin{lstlisting}[style=customxml]
<?xml version="1.0" encoding="UTF-8" ?>
<psml version="1.1" energy_unit="hartree" length_unit="bohr"
 uuid="904272c0-e496-11e6-496d-7f30ff5a9a4e"
 xmlns="http://esl.cecam.org/PSML/ns/1.1">
 
<provenance creator="ATM4.1.3" date="23-JUN-17">
<annotation action="semilocal-potential-generation">
<input-file name="INP">
<![CDATA[#
#  PS generation with core corrections
#  GGA (Perdew-Burke-Ernzerhof) XC , relativistic
#  3d7 4s1 configuration
#
   pe   Fe, GGA, rcore=0.70
        tm2      3.0
 n=Fe c=pbr
       0.0       0.0       0.0       0.0       0.0       0.0
    5    4
    4    0      1.00      0.00    # 4s1
    4    1      0.00      0.00    # 4p0
    3    2      7.00      0.00    # 3d7
    4    3      0.00      0.00    # 4f0
      2.00      2.00      2.00      2.00      0.00      0.70
#                                                        |
#                                       Radius of pseudocore
]]>
</input-file>
</provenance>
\end{lstlisting}

    \label{table:provenance}
 \end{table}

\subsection{Pseudo-atom specification}

The \tag{pseudo-atom-spec} element contains basic information about
the chemical element, the generation configuration, and the type of
calculation.

Element identification is in principle straightforward, with either
the atomic number or the chemical symbol. However, in the interest of
generality, and to cover special cases, such as synthetic (alchemical)
atoms, the
chemical symbol can be an arbitrary label, given by the attribute
\att{atomic-label} (with the convention that when possible the first
two characters give the standard chemical symbol) and the atomic
number a real number instead of a simple integer, given by the
attribute \att{atomic-number}.
 
 The atomic calculation can be done non-relativistically, with
 scalar-relativistic corrections (i.e., with the mass and Darwin
 terms)~\cite{Koelling-77,Takeda-78}, or with the full Dirac equation,
 including spin-orbit effects. In the latter case, the
 (semilocal) pseudopotentials are typically provided in two sets: the
 degeneracy-averaged $j=l+1/2$ and $j=l-1/2$ components, appropriate
 for scalar-relativistic use, and the spin-orbit
 components~\cite{Bachelet-82}.

 It is also possible to carry out a non-relativistic calculation for a
 spin-polarized reference configuration (using spin-DFT). Standard
 practice is then to keep only the population-averaged
 pseudopotentials, as a closed-shell frozen core should not be
 represented by a spin-dependent potential. (But see
 Ref.~\cite{Watson-98} for a different point of view.)

 In practice, then, one might have a primary set of
 $l$-dependent $V_l(r)$ potentials, possibly the result of averaging,
 and, in the case of fully-relativistic calculations, a spin-orbit
 set. In other cases, the generation program might output the original
 $lj$ versions in the Dirac case, or the ``up'' and ``down''
 components for a spin-DFT calculation.

 We cover all these possibilities in {\sc psml} by using the
 attribute \att{relativity}, with possible values ``no'',
 ``scalar'' or ``dirac'', and the optional attribute \att{spin-dft} to indicate
 (with value ``yes'') a non-relativistic spin-DFT calculation. Further, as detailed
 below, we provide support for the possible presence of various \textsl{sets}
 of magnitudes.

The pseudopotential construction is fundamentally dependent on the
core-valence split. In most cases it is clear which states are to be
considered as ``valence'', and which ones are to be kept in the frozen
core. However, there are borderline cases in which one has the option
to treat as ``valence'' so-called ``semicore'' states which are
relatively shallow and/or exhibit a sizable overlap with proper
valence states.

This information is provided in the mandatory
\tag{valence-configuration} element, which details the valence
configuration used at the time of pseudopotential generation, given by
the $n$ and $l$ quantum numbers, and the electronic occupation of each
shell.  Empty shells can be omitted.  In spin-DFT calculations,
the spin-up and spin-down occupations are also given.  This element
has a mandatory attribute \att{total-valence-charge} which contains
the total integrated valence charge $Q_{\rm val}$ for the
configuration used to generate the pseudopotential.

The core configuration can be determined easily from the knowledge of
the valence shells, but for completeness it can be given in the
optional \tag{core-configuration} element, with the same structure.
This tag would be useful, for example, if a core-hole
pseudopotential has been created~\cite{Garcia-Gil-12}.

The difference between the number of protons in the nucleus and the
sum of the populations of the core shells is the effective atomic
number of the pseudo-atom $Z_{\rm pseudo}$, which must be given in the
mandatory attribute \att{z-pseudo}.

The ``pseudization flavor'' or, more properly, a succint identifier
for the procedure for pseudopotential generation, is encoded in the
optional \att{flavor} attribute. If not present, more specific values
can be given per pseudopotential block and per pseudization channel
(see below).

If non-linear core-corrections~\cite{Louie-82} are present, the
optional \att{core-corrections} attribute must be set to ``yes''.

One further piece of information is needed to complete the
general specification of the computational framework: the type of
exchange-correlation (XC) functional(s) used. With the explosive growth
in the number of functionals, it is imperative that a robust naming
convention be used. In the absence of a general registry of
universally-agreed names, we propose a dual naming scheme. 
The element \tag{exchange-correlation} contains:
\begin{itemize}
\item A mandatory element \tag{libxc-info} that maps whatever XC
  functional is used in the generation code to the standard set of
  functionals in the {\sc Libxc} library~\cite{Marques-12,libxc}.
  This library supports
  a large number of functionals, and new ones are added promptly as
  their details are published. We thus require that
  pseudopotential-generation programs producing {\sc psml} files, and programs using
  the {\sc psml} format, provide internal tables mapping any built-in XC
  naming schemes to the {\sc Libxc} one. This element has the attribute
  \att{number-of-functionals}, and as many \tag{functional} elements
  as indicated by this attribute, with attributes \att{name},
  and \att{id},  as shown in the example
  of Table~\ref{table:pseudo-atom-spec}. These correspond to the 
  {\sc Libxc} identification
  standard. The attribute \att{type}
  (with values ``exchange'', ``correlation'',
  or ``exchange-correlation'') is optional. 
  Further, to support arbitrary mixtures of functionals, the
  optional attribute \att{weight} can also be indicated. 
\item An (optional) element \tag{annotation} (see
  Sec.~\ref{sec:annotations}) that can contain any XC
  identification used by the creator of the file, in the form of
  attribute-value pairs. This information can be read in an ad-hoc
  fashion by client programs, but it is obviously not as complete or
  robust as that contained in the \tag{libxc-info}. For maximum
  interoperability, client programs should thus implement an interface
  to {\sc Libxc}.
\end{itemize}
 \begin{table}[h]
   \caption{ An example of the \tag{pseudo-atom-spec} element }
   \begin{lstlisting}[style=customxml]
<pseudo-atom-spec atomic-label="Se" z-pseudo="6" atomic-number="34"
 flavor="Troullier-Martins" relativity="dirac" spin-dft="no"
 core-corrections="yes">
<exchange-correlation>
<annotation oncvpsp-xc-code="3"
 oncvpsp-xc-type="LDA -- Ceperley-Alder Perdew-Zunger" />
<libxc-info number-of-functionals="2">
<functional name="Slater exchange (LDA)" type="exchange" id="1" />
<functional name="Perdew &amp; Zunger (LDA)" type="correlation" id="9" />
</libxc-info>
</exchange-correlation>
<valence-configuration total-valence-charge="6">
<shell n="4" l="s" occupation="2" />
<shell n="4" l="p" occupation="4" />
</valence-configuration>
</pseudo-atom-spec>
\end{lstlisting}
     \label{table:pseudo-atom-spec}
 \end{table}

An optional \tag{annotation} element can also be included inside the
\tag{pseudo-atom-spec} element. Note that the ordering of child
elements is significant (see Sec.~\ref{sec:schema}).

\subsection{Radial functions and grid specification}

At the core of this new format we face the problem of how to store a
variety of different radial functions (semilocal pseudopotentials,
projectors, pseudo wave-functions, pseudocore and valence
charges, etc.) in a radial grid. Most pseudopotential-generation codes
export their data for their radial functions $f(r)$ as a tabulation
$\{f(r_i)\}$, where $\{r_i\}$ are discrete values of the radial
coordinate in an appropriate mesh.

A variety of meshes with different functional forms and
parametrization details are in common use.  Our preferred way to
handle this variety of choices is to specify the actual grid point
data in the file. This is most extensible to any kind of grid, and
avoids problems of interpretation of the parameters, starting and
ending points, etc. Furthermore, as explained below, the {\sc psml} handling
library is completely grid-agnostic, since evaluators are provided for
the relevant functions $f(r)$, in such a way that a client code can
obtain the value of $f$ at any radial coordinate $r$, in particular at
the points of a grid of its own choosing. The precision of the
computed value $f(r)$ is of course dependent on the quality of the
$f(r_i)$ tabulation in the first place, and producer codes should take
this issue seriously. We discuss more points related to the evaluation
of tabulated data in Sec.~\ref{sec:interpolation}.

The format should be flexible enough to allow each radial function to
use its own grid if needed, while providing for the most common case
in which all radial functions use a common grid (or a subset of its
points).  Our solution is to encode the information about each radial
function in a \tag{radfunc} element, which contains the tabulation
data $f(r_i)$ in a \tag{data} element. The grid specification uses a
cascade scheme with an (optional but recommended) top-level \tag{grid}
element, optional mid-level \tag{grid} elements under certain grouping
elements, and at the lowest level optional \tag{grid} elements inside
the individual \tag{radfunc} elements.  The grid $\{r_i\}$ for a function
is inherited from the closest \tag{grid} element at an enclosing level
if it is not specified in the local \tag{radfunc} element. The
grouping elements currently allowed to include mid-level grids are
those for semilocal potentials, nonlocal projectors, local potential,
pseudo-wavefunctions, valence charge, and pseudocore charge.

The \tag{grid} elements should have a mandatory ``npts'' attribute
providing the number of points, and a \tag{grid-data} element with the
grid point data as formatted real numbers with appropriate
precision. All radial data must be given in bohr.

For convenience, it is allowed to include an \tag{annotation} element
as child of \tag{grid}, with appropriate attributes, to provide
additional information regarding the form of the grid data. The client
program can process this information if needed.
The \tag{data} element may contain an optional attribute \att{npts} to
indicate the number of values that follow. In its absence, the number
of values must match the size of the grid. The \att{npts} attribute is
useful in those cases in which the effective range of a radial
function is significatively smaller than the extent of the
grid. For example, a {\sc psml} file might contain a
top-level grid with a large range, appropriate for the valence charge
density, of which only a subset of points are used for the projectors,
which have a much smaller range.

A technical point should be kept in mind. When processing a {\sc psml} file,
the radial function information is typically stored internally as a
table on which interpolation is performed to obtain values of the
function at specific radii. In order to avoid the dangers associated
with extrapolation, the radial grid must contain as first point $r=0$,
and any radial magnitudes (pseudopotentials, wave-functions, pseudo-core or
valence charges) should be given without extra factors of $r$ that
might hamper the calculation of needed values at $r=0$. In this
way the processor can unambiguously determine the function values at
all radial points.

When evaluating a function at a point $r$ beyond the maximum range of
the tabulated data in the {\sc psml} file, a processor should return:
\begin{itemize}
\item -$Z_{\rm pseudo}/r$ for the semi-local and local
  pseudopotentials, in keeping with the well-known asymptotic
  behavior.
\item Zero for projectors, pseudo-wave-functions and valence and pseudo-core
  charges.
\end{itemize}
 \begin{table}[h]
    \caption{ An example of the \tag{grid} element }
   \begin{lstlisting}[style=customxml]
<grid npts="1186">
<annotation type="log-atom" nrval="1186" scale="0.442634317262E-04"
 step="0.125000000000E-01" />
<grid-data>
  0.000000000000E+00  5.567654309900E-07  1.120534108973E-06  1.691394123953E-06
  2.269434673967E-06  2.854746079028E-06  3.447419795234E-06  4.047548429058E-06
  ...
</grid-data>
</grid>
\end{lstlisting}
    \label{table:grid}
 \end{table}

\subsection{Semilocal components of the pseudopotentials}
\label{sec:semilocal}

When available, the $l$-dependent (or maybe $lj$-dependent) semilocal
components $V_l(r)$ of the pseudopotential are classified under
\tag{semilocal-potentials} elements, with attributes:
              
 \begin{itemize}
    \item \att{set}: A string indicating which \textsl{set} (see below) the potentials
      belong to. If missing, the information is obtained from the
      records for the individual potentials.

    \item \att{flavor}: (optional) The pseudization flavor. If
      missing, its value is inherited from the value in the
      \tag{pseudo-atom-spec} element. It can also be superseded by the
      records for the individual potentials.
 \end{itemize}

The \att{set} attribute allows the handling of various sets of
pseudopotentials. Its value is normalized as follows, depending on the
type of calculation generating the pseudopotential and the way in
which the code chooses to present the results:
 \begin{itemize}
 \item ``non\_relativistic'' for the non-relativistic, non-spin-DFT
   case.
 \item ``scalar\_relativistic'' if the calculation is
   scalar-relativistic, or if it is fully relativistic and an
   set of $lj$ potentials averaged over $j$ is provided.
 \item ``spin\_orbit'' if a fully relativistic code provides this
   combination of $lj$ potentials.
 \item ``lj'' for a fully relativistic calculation with straight
   output of the $lj$ channels.
 \item ``spin\_average'' for the spin-DFT case when the
   generation code outputs a population-averaged pseudopotential.
 \item ``up'' and ``down'',  for a spin-DFT calculation with straight
   output of the spin channels.
 \item ``spin\_difference'' for the spin-DFT case when the
   generation code outputs also the difference between the ``up'' and
   ``down'' potentials. This and the previous case are retained for
   historical reasons, but are likely used rarely.
\end{itemize}

Note that a given code might choose to output its semilocal-potential
information in two different forms (say, as scalar-relativistic and
spin-orbit combinations plus the $lj$ form). The format allows this,
although in this particular case the information can easily be
converted from the $lj$ form to the other by client programs.

For extensibility, the format allows two more values for the \att{set}
attribute, ``user\_extension1'' and ``user\_extension2'', which can
in principle be used to store custom information while maintaining
structural and operative compatibility with the format.

The pseudopotentials must be given in hartree.

The \tag{semilocal-potentials} element contains child \tag{slps} elements, 
which store the information for the individual semilocal
pseudopotential components.  The attributes of this element are:

 \begin{itemize}
    \item \att{n}: principal quantum number of the pseudized shell.
    \item \att{l}: angular momentum number of the pseudized shell.
    \item \att{j}: (compulsory for ``lj'' sets) $j$ quantum number
    \item \att{rc}: $r_c$ pseudization radius for this shell
          (in Bohr).
    \item \att{eref}: (optional) reference energy (eigenvalue) of
      the all-electron wavefunction to be pseudized (in
      hartree).
    \item \att{flavor}: (optional) To allow for different schemes for
      different channels, the value of this attribute, when present,
      takes precedence over the \att{flavor} attributes in the
      \tag{pseudo-atom-spec} element and the
      \tag{semilocal-potentials} elements.
 \end{itemize}

The optional attribute \att{eref} might only be meaningful for
certain potentials (for example, those that have been directly
generated, and are not the product of any extra conversion, such
as from $lj$ to scalar-relativistic plus spin-orbit form).

Each \tag{slps} element contains a \tag{radfunc}
element. The order in which the \tag{slps} elements appear
is irrelevant.

The \tag{semilocal-potentials} element can contain an optional \tag{grid}
child applying to all the enclosed elements, as well as an optional
\tag{annotation} element for arbitrary extra information in key-value format.

\subsection{Pseudopotential in fully non-local form. }

Most modern electronic-structure codes do not actually use the
pseudopotential in its semi-local form, but in a more efficient fully
non-local form based on short-range projectors plus a ``local''
potential:
\begin{equation}
\hat{V}_{ps}  = \hat{V}_{\rm local} + \sum_i {|\chi_i> E_{\rm KB}^i
  <\chi_i|}
\label{eq:fully-non-local}
\end{equation}
proposed originally by Kleinman and Bylander~\cite{Kleinman-82} and generalized
among others by Bl{\"o}chl~\cite{Bloch-94}, Vanderbilt~\cite{Vanderbilt-90}, 
and Hamann~\cite{Hamann-13}. 

The information about this operator form of the pseudopotential is
split in two elements, holding the local potential and the nonlocal projectors.

\subsubsection{Local potential}

The \tag{local-potential} element has the attribute

 \begin{itemize}
    \item \att{type} We suggest the string ``l=X'' when the local
      potential is taken to be the semi-local component for channel
      ``X'', or any other succint comment if not.
 \end{itemize}

and contains a \tag{radfunc} element with the actual data for
the local pseudopotential.

Optionally, the \tag{local-potential} element can contain a child
\tag{local-charge} element, describing a radial function $\rho_{\rm
  local}(r)$ related to ${V}_{\rm local}(r)$ by Poisson's
equation. That is, $\rho_{\rm local}(r)$, integrating to $Z_{\rm
  pseudo}$ and localized in the core region, is the effective charge
that would generate ${V}_{\rm local}(r)$. The \tag{local-charge} element is
optional because not all ${V}_{\rm local}(r)$ functions are
representable as originating from a charge distribution
(${V'}_{\rm
  local}(0)$ must be zero for this). When it is present, however, it
can save some client programs (such as {\sc siesta}, which uses $\rho_{\rm
  local}(r)$ to generate a very convenient localized neutral-atom
potential) the task of computing numerical derivatives of ${V}_{\rm
  local}(r)$.

The \tag{local-potential} element can also contain an optional \tag{grid}
child applying to all the enclosed elements, as well as an optional
\tag{annotation} element for arbitrary extra information in key-value format.

\subsubsection{Non-local projectors}

The information about the non-local projectors is stored in
\tag{nonlocal-projectors} elements, with the optional attribute
\att{set}, as above, containing \tag{proj} elements with attributes

 \begin{itemize}
    \item \att{ekb}: Prefactor of the projector in the corresponding
      term in Eq.~\ref{eq:fully-non-local} (in hartree).
    \item \att{eref}: (optional) Reference energy used in the
      generation of the projector (in hartree).
    \item \att{l}: angular momentum number 
    \item \att{j}: (compulsory for ``lj'' sets) $j$ quantum number
    \item \att{seq}: sequence number within a given $l$ (or $lj$)
      shell, to support the case of multiple projectors.
    \item \att{type}: Succint comment about the kind of projector
 \end{itemize}

and a \tag{radfunc} element containing the data for the $\chi_i$
functions in Eq.~\ref{eq:fully-non-local}. These functions are formally
three-dimensional, including the appropriate spherical harmonic for
the angular coordinates, and a radial component: $\chi_i=
\chi_i(r)Y_{lm}(\theta,\phi)$. What is actually stored in the file is
the function $\chi_i(r)$, normalized in the one-dimensional sense:
\begin{equation}
        \int_{0}^{\infty} r^{2} \left| \chi_i(r) \right|^{2} dr=
        \int_{0}^{\infty} \left| r\chi_i(r) \right|^{2} = 1,
        \label{eq:normkb}
\end{equation}
and proportional to $r^{l}$ near the origin.

There can be several \tag{nonlocal-projectors} elements, using
different values for the \att{set} attribute, as explained in
Sec.~\ref{sec:semilocal}.  For projectors in the ``dirac'' case, the
functionality provided by the handling of \att{set} attributes can be
very useful, as it is not straightforward to convert the $lj$
information into scalar-relativistic and spin-orbit combinations.
Unlike in the semi-local case, this conversion is not reversible, so
the $lj$ form is more fundamental.  For maximum interoperability,
producer codes should store both the $lj$ and the combination sets.

The optional attribute \att{eref} might only be meaningful for
certain projectors (for example, those that have been directly
generated, and are not the product of any extra conversion, such
as from $lj$ to scalar-relativistic plus spin-orbit form).

\subsection{Pseudo-wave functions}
\label{sec:pseudo-wave-functions}

Pseudo-wavefunctions are typically produced at an intermediate stage in the
generation (and testing) of a pseudopotential, but they are not
strictly needed in electronic-structure codes, except in a few cases:
\begin{itemize}
\item When atomic-like initial wavefunctions are needed to start the
  electronic-structure calculation.
\item When the code uses internally a fully-nonlocal form of the
  pseudopotential which is constructed from the semilocal form and the
  pseudo-wavefunctions.
\end{itemize}
  
While these pseudo-wavefunctions could be generated by the client
program, we allow for the possibility of including them explicitly in
the {\sc psml} file.
Any (optional) wavefunction data must be included in
\tag{pseudo-wave-functions} elements, with a \att{set} attribute and
as much extra metadata as needed (which we do not try to standardize
at this point, so it should be given in the form of annotations).

The extra metadata might indicate whether the data is for actual pseudized
wave-functions, or for the pseudo-valence wave functions generated with
the obtained pseudopotential. There might be subtle differences
between them, notably regarding relativistic effects, as some
generation codes use a non-relativistic scheme to ``test'' the
pseudopotential and generate pseudo wavefunctions, instead of a scalar
or fully relativistic version.

Each pswf is given in a \tag{pswf} element, with attributes that
identify the quantum numbers for the shell and the apropriate energy
level (which could be the eigenvalue in the original pseudization, or
another energy level used in the integration leading to the wavefunction):

 \begin{itemize}
    \item \att{n}: principal quantum number of the shell. 
    \item \att{l}: angular momentum number of the shell.
    \item \att{j}: (compulsory for ``lj'' sets) $j$ quantum number
    \item \att{energy\_level}
 \end{itemize}

and a \tag{radfunc} element.

The data is for the standard radial part of the wave function
$R_{n,l}(r)$, and (for bound states) should be normalized as
      \begin{equation}
        \int_{0}^{\infty} r^{2} \left| R_{n,l}(r) \right|^{2} dr=
        \int_{0}^{\infty} \left| u_{n,l}(r) \right|^{2} = 1.
        \label{eq:normpswf}
      \end{equation}
Within our {\sc psml} format $R(r)$ is given, rather than $u(r)$, 
due to the extrapolation issues detailed in the section covering the grid.

The \tag{pseudo-wave-functions} elements can contain an optional \tag{grid}
child applying to all the enclosed elements.

\subsection{Valence charge density}
\label{sec:valence-charge}

Some electronic-structure codes might need information about the
valence charge density of the pseudo-atom. This can be 
the pseudo-valence charge density used to unscreen
the ionic potential during the pseudopotential generation process, or
the pseudo-valence charge computed from the pseudopotential itself. In {\sc psml}
it is given under the \tag{valence-charge} element, whose
child \tag{radfunc} element holds a solid-angle-integrated 
form $q(r)$ normalized so that:
 \begin{equation}
    \int_{0}^{\infty} r^2 q(r) dr = Q.
    \label{eq:normrhoval}
 \end{equation}
Here $Q$ is the total charge output, which must be stored in the
\att{total-charge} attribute. The attributes
\att{is-unscreening-charge} (with value ``yes'' or ``no'') and
\att{rescaled-to-z-pseudo} (``yes'' or ``no'') are 
optional. 

In combination with the information in the \tag{valence-configuration}
element, these attributes will help some client codes process the
valence charge density data appropriately, particularly in the case in
which the pseudopotential generation used an ionic configuration. Some
codes output in this case the unscreening charge rescaled to
$Z_{\rm pseudo}$.
The \tag{valence-charge} can also contain an optional \att{annotation}
element.

\subsection{Pseudocore-charge density}
\label{sec:core-charge}

An (optional) smoothed charge density matching the density of the core
electrons beyond a certain radius, for use with a non-linear core
correction scheme~\cite{Louie-82}, is contained within the
\tag{pseudocore-charge} element, with the data in the same
solid-angle-integrated form (and implicit units) as the valence
charge, and with the extra optional attributes:
\begin{itemize}
\item \att{matching-radius}: The point $r_{\rm core}$ at which the true core
  density is matched to the pseudo-core density.
\item \att{number-of-continuous-derivatives}: In the original scheme
  by Louie {\it et al.}~\cite{Louie-82} the pseudo-core charge was
  represented by a two-parameter formula, providing continuity of the
  first derivative only. Other typical schemes provide continuity of
  the second and even higher derivatives. It is expected that the
  number of continuous derivatives, rather than the detailed form of
  the matching, is of more interest to a client program.
\end{itemize}

An optional \att{annotation} element with appropriate attributes might
be given to document any extra details of the model-core generation.
The \tag{pseudocore-charge} element must appear if the
\att{core-corrections} attribute of the \tag{pseudo-atom-spec} element has the
value ``yes''.

\subsection{The handling of annotations}
\label{sec:annotations}

XML provides for built-in extensibility and client programs can use as
much or as little information as needed. For the actual mapping of a
domain ontology to a XML-based format, however, clients and producers
have to agree on the terms used. What has been described in the above
sections is a minimal form of such a mapping, containing the basic
concepts and functions needed. The extension of the format with new
fixed-meaning elements and attributes would involve an updated schema
and re-coding of parsers and other programs. A more light-weight
solution to the extensibility issue is provided by the use of
\textsl{annotations}, which have the morphology of XML empty elements
(containing only attributes) but can appear in various places and
contain arbitrary key-value pairs. Annotations provide immediate
information to human readers of the {\sc psml} files, and can be
exploited informally by client programs to extract additional
information. For the latter use, it is clear that some degree of
permanence and agreed meaning should be given to annotations, but this
task falls not on some central authority, but on specific codes.

Annotations are currently allowed within the following elements:
\tag{provenance},
\tag{pseudo-atom-spec},
\tag{exchange-correlation},
\tag{valence-configuration},
\tag{core-configuration},
\tag{semilocal-potentials},
\tag{local-potential},
\tag{nonlocal-projectors},
\tag{pseudo-wave-functions},
\tag{valence-charge},
\tag{core-charge}, and
\tag{grid}.
Top-level annotations are not allowed. They properly
belong in the \tag{provenance} elements.

\section{Formal specification of the {\sc psml} format}
\label{sec:schema}

We provide a formal XML schema for the {\sc psml} format, given in the
very readable RELAX NG  compact form~\cite{relax-ng}. This schema
can be used directly for validation of {\sc psml} files,  or converted to an
W3C schema file (using respectively the \texttt{jing} and
\texttt{trang} tools of the RELAX NG project).

This is the overall structure of a {\sc psml} document, showing the main
building blocks. 
The definitions of the grammar elements are given in
the next section, when the closely related API is discussed.

\begin{lstlisting}[style=customrnc]
default namespace = "http://esl.cecam.org/PSML/ns/1.1"

PSML =  element psml { 
            Root.Attributes         
          , Provenance+             # One or more provenance elements
          , PseudoAtomSpec          
          , Grid?                   # Optional top-level grid
          , ValenceCharge           
          , CoreCharge?             # Optional pseudo-core charge
		, (
		     (SemiLocalPotentials+ , PSOperator?)   
		     |
		     (SemiLocalPotentials* , PSOperator )
		  )
          , PseudoWaveFunctions*          # Zero or more Pseudo Wavefunction groups
       }  

PseudoAtomSpec =  element pseudo-atom-spec {    PseudoAtomSpec.Attributes
                                              , Annotation?
                                              , ExchangeCorrelation
		                              , ValenceConfiguration
                                              , CoreConfiguration?
                                           }

PSOperator =   (   LocalPotential            # Local potential
                 , NonLocalProjectors* )     # Zero or more fully nonlocal groups
\end{lstlisting}

The quantifiers '*', '+', and '?' mean ``zero or more'', ``one or
more'', and ``at most one'', and the '$\mid$' sign expresses an exclusive
``or'' (choice) operation. Even though RELAX NG can accommodate
interleaved elements, this feature is not fully representable in W3C
schema, and the ordering of the elements above is strict.
We use a URI in the ``esl.cecam.org'' domain to identify
the schema namespace, but note that it is not a resolvable location.

The main feature not discussed above in Sec.~\ref{sec:psml-format} is
the constraint of \textsl{non-emptiness} of the {\sc psml} file: it must contain
at least either a set of semilocal-potentials, or a complete
pseudopotential operator consisting of a local potential and a set of
nonlocal projectors. It is also possible to have a single local
potential as a degenerate form of pseudopotential operator. Beyond the
minimal requirements, a {\sc psml} file can contain multiple occurrences of
any of these elements.

\section{The {\sc psml} library}
\label{sec:libpsml}
We provide a companion library to the {\sc psml} format, \libpsml{}, that
provides transparent parsing of and data extraction from {\sc psml} files, as
well as basic editing and data conversion capabilities.

The library is built around a data structure of type \textsl{ps\_t}
that maps the information in a {\sc psml} file. Instances of this
structure are populated by {\sc psml} parsers, processed by
intermediate utility programs, and used as handles for information
retrieval by client codes through accessor routines. The library
provides, in essence:
\begin{enumerate}[(i)]
\item  A routine to parse a {\sc psml} file and produce a \textsl{ps}
  object of type \textsl{ps\_t}.
\item  A routine to dump the information in a \textsl{ps} object to a
  {\sc psml} file.
\item  Accessor routines to extract information from \textsl{ps}
  objects.
\item  Some setter routines to insert specific blocks of information
  into \textsl{ps} objects. These might be used by intermediate
  processors or by high-level parsers.
\end{enumerate}

The library is written in modern Fortran and provides a
high-level Fortran interface. A C/C++ interface is in preparation.

An example of use of the library is provided in
Table~\ref{table:data-access}.

In what follows we describe the basic exported data structures and
procedures. Full documentation for the library, as well as general
information about the {\sc psml} format ecosystem, is available at the
{\sc psml} reference page under the Electronic Structure Library
project website: \textsl{http://esl.cecam.org/PSML}.

 \begin{table}[h]
    \caption[ ]{An example of the idioms used in the \libpsml{} library}
 \rule{\textwidth}{0.4pt}
\begin{lstlisting}[style=customfortran]
use m_psml
type(ps_t)   :: ps

call psml_reader(filename,ps)

! Set up our grid
npts = 400; delta = 0.01  
allocate(r(npts))
do ir = 1, npts
   r(ir) = (ir-1)*delta
enddo

call ps_Potential_Filter(ps,set=SET_SREL,indexes=idx,number=npots)
do i = 1, npots
   call ps_Potential_Get(ps,idx(i),l=l,n=n,rc=rc)
   ...
   do ir = 1, npts
      val = ps_Potential_Value(ps,idx(i),r(ir))
      ...
   enddo
enddo
\end{lstlisting}
      \rule{\textwidth}{0.4pt}
    \label{table:data-access}
 \end{table}

\subsection{Exported types}
\label{sec:exported-types}

The library exports a few fortran derived types to represent opaque
handles in the routines:
\begin{itemize}
\item \name{ps\_t}

  This is the type for the handle which should be passed to most routines in
  the API
\item \name{ps\_annotation\_t}

  The associated handle is used in the routines that create annotations or
  extract the data in them (see Section~\ref{sec:annotation-api}).
  
\item \name{ps\_radfunc\_t}

  This corresponds to the internal implementation of a radial 
  function. Its use is mostly reserved for low-level operations, as
  the API provides convenience evaluators for most functions.
  
\end{itemize}

Additionally, and to avoid ambiguities in real types, the library
exports the integer parameter \name{ps\_real\_kind} that represents the \textsl{kind} of
the real numbers accepted and returned by the library.
  
\subsection{Parsing}

\begin{itemize}
\item \routine{psml\_reader} (\arg{filename, ps, debug})

parses the {\sc psml} file \arg{filename} and populates the data
structures in the handle \arg{ps}. An optional
\arg{debug} argument determines whether the library issues
debugging messages while parsing.

\item \routine{ps\_destroy} (\arg{ps})

is a low-level routine provided for completeness in cases where a
pristine \arg{ps} is needed for further use.

\end{itemize}

\subsection{Library identification}

\begin{itemize}
\item
 \arg{function} \routine{ps\_GetLibPSMLVersion}()
 \arg{result}(\arg{version})

    The version is returned as an integer with the two least
    significant digits associated to the patch level (for example:
    1106 would correspond to the typical dot form 1.1.6).
    
\end{itemize}

\subsection{Data accessors}

The API follows closely the element structure of the {\sc psml}
format. Each section in the high-level document structure of
Sec.~\ref{sec:schema} is mapped to a group of routines in the
API. Within each, there are routines to query any internal structure
(attributes, existence, number, or selection of child elements) and
routines to obtain specific data items (attributes, content of child
elements).

\subsubsection{Root attributes}
\begin{lstlisting}[style=customrnc]
Root.Attributes =  attribute energy_unit { "hartree" }
                 , attribute length_unit { "bohr" }
                 , attribute uuid { xsd:NMTOKEN }
                 , attribute version { xsd:decimal }
\end{lstlisting}

\begin{itemize}
\item \routine{ps\_RootAttributes\_Get}
  (\arg{ps,uuid,version,namespace})

  As in all the routines that follow, the handle \arg{ps} is
  mandatory. All other arguments are optional, with ``out'' intent,
  and of type \name{character(len=*)}. 
  \arg{version} returns the {\sc psml} version of the file being
  processed. A given version of the library is able to process files
  with lower version numbers, up to a reasonable limit. For our purposes,
  the \name{NMTOKEN} specification refers to a string without spaces or commas.
  
\end{itemize}

\subsubsection{Provenance data}
\begin{lstlisting}[style=customrnc]
Provenance =  element provenance {
                attribute record-number { xsd:positiveInteger }?
              , attribute creator { xsd:string }
              , attribute date { xsd:string }
      
              , Annotation?
	      , InputFile*    # zero or more input files
              }
	      
InputFile =  element input-file {
               attribute name { xsd:NMTOKEN }, # No spaces or commas allowed
               text
             }
\end{lstlisting}

As there can be several \tag{provenance} elements, the API provides a
function to enquire about their number (\arg{depth} of provenance
information), and a routine to get the information from a given level:

\begin{itemize}

\item \arg{function} \routine{ps\_Provenance\_Depth}(\arg{ps})  \arg{result}(\arg{depth})

  The argument \arg{depth} returns an integer number.
  
\item
  \routine{ps\_Provenance\_Get}(\arg{ps,level,creator,date,annotation,number\_of\_input\_files})

  The integer argument \arg{level} selects the provenance depth level
  (1 is the deepest, or older, so to get the latest record the routine
  should be called with \arg{level}=\arg{depth} as returned from the
  previous routine). All other arguments are optional with ``out''
  intent. \arg{creator} and \arg{date} are strings. Here and in what
  follows, \arg{annotation} arguments are of the opaque type
  \textsf{ps\_annotation\_t} (see Section~\ref{sec:exported-types}).
  If there is no annotation, an empty structure is returned. The
  information in an annotation object can be accessed using routines
  described in Section~\ref{sec:annotation-api}.
  
\end{itemize}

\subsubsection{Pseudo-atom specification attributes and annotation}

\begin{lstlisting}[style=customrnc]
PseudoAtomSpec.Attributes = 
       attribute atomic-label { xsd:NMTOKEN },
       attribute atomic-number { xsd:double },
       attribute z-pseudo { xsd:double },
       attribute core-corrections { "yes" | "no" },
       attribute relativity { "no" | "scalar" | "dirac" },
       attribute spin-dft { "yes" | "no" }?,
       attribute flavor { xsd:string }?
\end{lstlisting}

\begin{itemize}
\item \routine{ps\_PseudoAtomSpec\_Get} (\arg{ps, atomic\_symbol,
  atomic\_label, atomic\_number, z\_pseudo,
  pseudo\_flavor, relativity, spin\_dft, core\_corrections, annotation})

  The arguments \arg{spin\_dft} and \arg{core\_corrections} are boolean,
  and the routine returns an empty string in \arg{flavor} if the
  attribute is not present (recall that \att{flavor} is a cascading
  attribute that can be set at multiple levels).
  The arguments \arg{atomic\_number} and \arg{z\_pseudo} are reals of
  kind \name{ps\_real\_kind} (see Sec.~\ref{sec:exported-types}).

\end{itemize}

\subsubsection{Valence configuration}
\begin{lstlisting}[style=customrnc]
ValenceConfiguration =  element valence-configuration {
                          attribute total-valence-charge { xsd:double },
	                  Annotation?,
                          ValenceShell+
                        }

ValenceShell =   Shell

Shell =  element shell {
           attribute_l,
           attribute_n,
           attribute occupation { xsd:double },
           attribute occupation-up { xsd:double }?,
           attribute occupation-down { xsd:double }?
         }

attribute_l = attribute l { "s" | "p" | "d" | "f" | "g" }
attribute_n = attribute n { "1" | "2" | "3" | "4" | "5" | "6" | "7" | "8" | "9" }
\end{lstlisting}

\begin{itemize}
\item
  \routine{ps\_ValenceConfiguration\_Get}(\arg{ps,nshells,charge,annotation})

  This routine returns (as always, in optional arguments), the
  values of the top-level attributes, any annotation, and the number of
  \texttt{Shell} elements, which serves as upper limit for the index
  \arg{i} in the following routine, which extracts shell information:
  
\item
  \routine{ps\_ValenceShell\_Get}(\arg{ps,i,n,l,occupation,occ\_up,occ\_down})

  The \arg{n} and \arg{l} quantum number arguments are integers
  (despite the use of spectroscopic symbols for the angular momentum
  in the format), and the occupations real.
  
\end{itemize}

\subsubsection{Exchange and correlation}
\begin{lstlisting}[style=customrnc]
ExchangeCorrelation =  element exchange-correlation {
                         Annotation?
                         , element libxc-info {
                              attribute number-of-functionals { xsd:positiveInteger },
                              LibxcFunctional+
                           }
                        }

LibxcFunctional =   element functional {
                      attribute id { xsd:positiveInteger },
                      attribute name { xsd:string },
                      attribute weight { xsd:double }?,

                      # allow canonical names and libxc-style symbols
	      
                      attribute type { "exchange" | "correlation" | "exchange-correlation" |
	                               "XC_EXCHANGE" | "XC_CORRELATION" |
		  	                  "XC_EXCHANGE_CORRELATION" }?
                    }
\end{lstlisting}

The routines follow the same structure as those in the previous section.

\begin{itemize}

\item
  \routine{ps\_ExchangeCorrelation\_Get}(\arg{ps,annotation,n\_libxc\_functionals})

\item
  \routine{ps\_LibxcFunctional\_Get}(\arg{ps,i,name,code,type,weight})

  The argument \arg{type} corresponds to the type of Libxc
  functional, and \arg{code} to the id number.
  
\end{itemize}

\subsubsection{Valence and Core Charges}
\begin{lstlisting}[style=customrnc]
ValenceCharge =  element valence-charge {
                   attribute total-charge { xsd:double },
                   attribute is-unscreening-charge { "yes" | "no" }?,
                   attribute rescaled-to-z-pseudo { "yes" | "no" }?,
                   Annotation?,
                   Radfunc
                 }  
    
# =========
CoreCharge =  element pseudocore-charge {
                attribute matching-radius { xsd:double },
                attribute number-of-continuous-derivatives { xsd:nonNegativeInteger },
                Annotation?,
                Radfunc
              }
\end{lstlisting}

These are radial functions with some metadata in the form of
attributes, an optional annotation, and a Radfunc child. The accessors
have the extra optional argument \arg{func} that returns a handle to
a \textsf{ps\_radfunc\_t} object, which can later be used to get extra
information.

\begin{itemize}
\item \routine{ps\_ValenceCharge\_Get}(\arg{ps,total\_charge,
                      is\_unscreening\_charge, rescaled\_to\_z\_pseudo,
                      annotation,func})

  The routine returns an emtpy string in \arg{is\_unscreening\_charge} and
  \arg{rescaled\_to\_z\_pseudo} if the attributes are not present in the
  {\sc psml} file. 
  
\item
  \routine{ps\_CoreCharge\_Get}(\arg{ps,rc,nderivs,annotation,func})

  \arg{rc} corresponds to the matching radius and \arg{nderivs} to the
  continuity information. Negative values are returned if the
  corresponding attributes are not present in the file.

\end{itemize}

The \arg{func} object can be used to evaluate the radial functions at
a particular point \arg{r}:

\begin{itemize}
\item function \routine{ps\_GetValue}(\arg{func,r}) result(\arg{val})
\end{itemize}

but the API offers some convenience functions
\begin{itemize}
\item function \routine{ps\_ValenceCharge\_Value}(\arg{ps,r}) result(\arg{val})
\item function \routine{ps\_CoreCharge\_Value}(\arg{ps,r}) result(\arg{val})
\end{itemize}

\subsubsection{Local Potential and Local Charge Density}

\begin{lstlisting}[style=customrnc]
LocalPotential =  element local-potential {
                    attribute type { xsd:string },
                    Annotation?,
                    Grid?,
                    Radfunc,
		    LocalCharge?  # Optional local-charge element
                  }
    
LocalCharge =  element local-charge {
	         Radfunc
	       }
\end{lstlisting}

\begin{itemize}
\item
  \routine{ps\_LocalPotential\_Get}(\arg{ps,type,annotation,func,has\_local\_charge,func\_local\_charge})

\end{itemize}

In this version of the API, the optional \tag{local-charge} element is
not given a first-class status. To evaluate it (if the boolean
argument \arg{has\_local\_charge} is true), the
\arg{func\_local\_charge} argument has to be used in the
\routine{ps\_GetValue} routine above. The local potential can be
evaluated via the \arg{func} object or with the convenience function 

\begin{itemize}
\item function \routine{ps\_LocalPotential\_Value}(\arg{ps,r}) result(\arg{val})
\end{itemize}

\subsubsection{Semilocal potentials}
\begin{lstlisting}[style=customrnc]
SemiLocalPotentials =  element semilocal-potentials {
                         attribute_set,
                         attribute flavor  { xsd:string }?,
                         Annotation?,
			 Grid?,
                         Potential+
                        }

Potential =   element slps {
                attribute flavor { xsd:string }?,
                attribute_l,
                attribute_j ?,
                attribute_n,
                attribute rc { xsd:double },
                attribute eref { xsd:double }?,
                Radfunc
              }   
\end{lstlisting}

As explained in Sec.~\ref{sec:psml-format}, there can be several
\tag{semilocal-potentials} elements corresponding to different
sets. Internally, the data is built up in linked lists during the
parsing stage and later all the data for the \tag{slps} child elements
are re-arranged into flat tables, which can be queried like a simple
database. The table indexes for the potentials with
specific quantum numbers, or set membership, can be obtained with
the routine

\begin{itemize}

\item \routine{ps\_SemilocalPotentials\_Filter}(\arg{ps,indexes\_in,l,j,n,set,indexes,number})

Here, the optional argument (of intent ``in'') \arg{indexes\_in} is an
integer array containing a set of indexes on which to
  perform the filtering operation. If not present, the full table is used.
The optional arguments \arg{l,j,n,set} are the values corresponding to
the filtering criteria. Upon return, the optional argument
\arg{indexes} would contain the set of indexes which match all the specified
criteria, and \arg{number} the total number of matches.

The \arg{set} argument has to be given using special integer symbols exported
by the API: \name{SET\_SREL, SET\_NONREL, SET\_SO, SET\_LJ, SET\_UP,
  SET\_DOWN, SET\_SPINAVE}, \name{SET\_SPINDIFF}, or the wildcard
specifier \name{SET\_ALL}.
An example of the use of this routine has already
been given in Table~\ref{table:data-access}.

\end{itemize}

The appropriate indexes can then be fed into the following routines to
get specific information:

\begin{itemize}
\item
  \routine{ps\_Potential\_Get}(\arg{ps,i,l,j,n,rc,eref,set,flavor,annotation,func})

  All arguments except \arg{ps} and \arg{i} are optional. The value
  returned in \arg{set} is an integer which can be converted to a
  mnemonic string through the \name{str\_of\_set} convenience function.
  The annotation returned corresponds to the optional \tag{annotation}
  element of the parent block of the \tag{slps} element.
  
  The routine returns a very large positive value in \arg{eref} if the
  corresponding attribute is not present in the file.

\item function \routine{ps\_Potential\_Value}(\arg{ps,i,r}) result(\arg{val})

\end{itemize}

\subsubsection{Nonlocal Projectors}
\begin{lstlisting}[style=customrnc]
NonLocalProjectors =  element nonlocal-projectors {
                        attribute_set,
                        Annotation?,
                        Grid?,
	                Projector+
                      }

Projector =   element proj {
                attribute ekb { xsd:double },
                attribute eref { xsd:double }?,
                attribute_l,
                attribute_j ?,
                attribute seq { xsd:positiveInteger },
                attribute type { xsd:string },
                Radfunc
              }+
\end{lstlisting}

The ideas are exactly the same as for the semilocal potentials. The
relevant routines are:

\begin{itemize}
\item  \routine{ps\_NonlocalProjectors\_Filter}(\arg{ps,indexes\_in,l,j,seq,set,indexes,number})
\item
  \routine{ps\_Projector\_Get}(\arg{ps,i,l,j,seq,set,ekb,eref,type,annotation,func})

  The routine returns a very large positive value in \arg{eref} if the
  corresponding attribute is not present in the file.
  
\item function \routine{ps\_Projector\_Value}(\arg{ps,i,r}) result(\arg{val})
\end{itemize}

\subsubsection{Pseudo Wavefunctions}
\begin{lstlisting}[style=customrnc]
PseudoWaveFunctions =  element pseudo-wave-functions {
                         attribute_set,
                         Annotation?,
                         Grid?,
                         PseudoWf+
                       }

PseudoWf =  element pswf {
       	      attribute_l,
              attribute_j ?,
              attribute_n,
              attribute energy_level { xsd:double} ?,
              Radfunc
            }
\end{lstlisting}

Again, the same strategy:

\begin{itemize}
\item \routine{ps\_PseudoWavefunctions\_Filter}(\arg{ps,indexes\_in,l,j,set,indexes,number})
\item  \routine{ps\_PseudoWf\_Get}(\arg{ps,i,l,j,n,set,energy\_level,annotation,func})

  The routine returns a very large positive value in \arg{energy\_level} if the
  corresponding attribute is not present in the file.

\item function \routine{ps\_PseudoWf\_Value}(\arg{ps,i,r}) result(\arg{val})
\end{itemize}

\subsection{Radial function and grid information}
\begin{lstlisting}[style=customrnc]
Radfunc =  element radfunc {
             Grid?,              # Optional grid element
             element data {      
               list { xsd:double+ }       # One or more floating point numbers
             }
           }

Grid =  element grid {
          attribute npts { xsd:positiveInteger },
          Annotation?,
          element grid-data {
                 list { xsd:double+ }   # One or more floating point numbers
          }
        }
\end{lstlisting}

In keeping with the {\sc psml} philosophy of being grid-agnostic, the basic
API tries to discourage the direct access to the data used in
the tabulation of the radial functions. The values of the functions at
a particular point \arg{r} can be generally obtained through the
\name{ps\_(Name)\_Value} interfaces, or through the \name{ps\_GetValue}
interface using \arg{func} objects of type \name{ps\_radfunc\_t}.

It is nevertheless possible to get annotation data for the grid of a
particular radial function, or for the top-level grid, through the
function

\begin{itemize}
\item function \routine{ps\_GridAnnotation}(\arg{ps,func}) result(\arg{annotation})

  If a radial function handle \arg{func} is given, the annotation for
  that radial function's grid is returned. Otherwise, the return value
  is the annotation for the top-level grid.  
\end{itemize}

\subsection{The evaluation engine}
\label{sec:interpolation}

In the current version of the library the evaluation of tabulated
functions is performed by default with polynomial interpolation, using
a slightly modified version of an algorithm borrowed (with permission)
from the {\sc oncvpsp} program by D.R. Hamann~\cite{oncvpsp}.  By
default seventh-order interpolation, as in {\sc oncvpsp}, is used.  If
the library is compiled with the appropriate pre-processor symbols,
the interpolator and/or its order can be chosen at runtime, but we
note that this should be considered a debugging
feature, as the reproducibility of results would be hampered if client codes
change the interpolation parameters at will. Generator codes should
instead strive to produce data tabulations that will guarantee a given
level of precision when interpolated with the default scheme, using
appropriate output grids on which to sample their internal data
sets. For example, our own work on enabling {\sc psml} output in {\sc
  oncvpsp} (see below) includes diagnostic tools to check the
interpolation accuracy.

Most codes use internally a non-uniform grid (e.g. logarithmic). We
have found that a good choice of output grid is a subset of the
producer's working grid points that leaves out most of the very close
points near the origin but maintains the rest. This can be achieved by
imposing a minimum inter-point separation $\delta$. This parameter
$\delta$ can be smaller than the typical linear-grid step used
currently by most codes, and still lead to smaller grids (in terms of
number of points) that preserve the accuracy of the output.

High-order interpolation can lead to \textsl{ringing} effects
(oscillations of the interpolating polynomial between points), notably
near edge regions when the shape of the function changes
abruptly. This is the case, for example, if the function drops to zero
within the interpolation range as a result of cutting off a tail. The
actual interpolated values will typically be very small, but might cause
undesirable effects in the client code. To avoid this problem, the
\libpsml{} evaluator works internally with an effective end-of-range
that is determined by analyzing the data values after parsing.

If needed for debugging purposes, the evaluator engine can be
configured by the routine:

\begin{itemize}
\item  \routine{ps\_SetEvaluatorOptions}(\arg{quality\_level,debug,
  use\_effective\_range,
  custom\_interpolator})

  All arguments are optional, and apply globally to the operation of
  the library. The \arg{custom\_interpolator} argument is not allowed
  if the underlying Fortran compiler does not support procedure
  pointers. \arg{quality\_level} (an integer) is by default and will
  typically be the interpolation order, but its meaning can change
  with the interpolator in use. The evaluator uses an effective range
  by default, as discussed above, but this feature can be turned off
  by setting \arg{use\_effective\_range} to \texttt{.false.}. The
  \arg{debug} argument will turn on any extra printing configured in
  the evaluator. By default, no extra printing is produced.
  
\end{itemize}

Finally, in case it is necessary to look at the raw tabular data for debugging 
purposes, the library also provides a low-level routine:

\begin{itemize}
\item \routine{ps\_GetRawData}(\arg{func,rg,data})

  It accepts a radial function handle \arg{func}, and the grid points
  and the actual tabulated data are returned in \arg{rg} and
  \arg{data}, which must be passed as allocatable real arrays.
  
\end{itemize}

\subsection{Editing of ps structures}

The {\sc psml} library has currently some limited support for editing the content
of \name{ps\_t} objects from user programs. For example, such an
editing might be done by a KB-projector generator to insert a new
provenance record (and KB and local-potential data) in the
\name{ps\_t} object, prior to dumping to a new {\sc psml} file.
Editing operations not yet supported directly by a given version can still
be carried out by a direct handling of the internal structure of the
\name{ps\_t} object, which is for now also visible to client
programs. 

\begin{itemize}
  \item\routine{ps\_RootAttributes\_Set}(\arg{ps,version,uuid,namespace})
  \item\routine{ps\_Provenance\_Add}(\arg{ps,creator,date,annotation})

  Annotations can be created by client programs using routines
  exported by the {\sc psml} API (see
  Section~\ref{sec:annotation-api}).
  
  \item\routine{ps\_NonlocalProjectors\_Delete}(\arg{ps})
  \item\routine{ps\_LocalPotential\_Delete}(\arg{ps})

  Only ``deletion'' operations are supported as yet.

\end{itemize}

\subsection{Dump of ps structures}

The contents of a (possibly edited) \name{ps\_t} object can be dumped
to a {\sc psml} file using the routine

\begin{itemize}
\item\routine{ps\_DumpToPSMLFile} (\arg{ps,fname,indent})

  Here \arg{fname} is the output file name, and \arg{indent} is a
  logical variable that determines whether automatic indenting of
  elements is turned on (by default it is not).
  
\end{itemize}

In principle, there could be \textsl{dumpers} for other file formats,
but their implementation is better left to specialized programs that
are clients of the library.
  
\subsection{Annotation API}
\label{sec:annotation-api}

To support the annotation functionality (see Sec.~\ref{sec:annotations}),
the library contains a module with a basic implementation of an
association list (a data structure holding key-value pairs). The library
exports the \name{ps\_annotation\_t} type, an empty annotation object
\name{EMPTY\_ANNOTATION}, and the following routines:

\begin{itemize}
\item\routine{reset\_annotation}(\arg{annotation})

  Cleans the contents of the \name{ps\_annotation\_t} object
  \arg{annotation} so that it can be reused.
  
\item\routine{insert\_annotation\_pair}(\arg{annotation,key,value,stat})

  Inserts the \arg{key, value} pair of character variables in the
  \name{ps\_annotation\_t} object \arg{annotation}. Internally,
  \arg{annotation} can grow as much as needed.
  
\item function \routine{nitems\_annotation}(\arg{annotation})
  result(\arg{nitems})

  Returns the number of key-value pairs in the annotation object

\item\routine{get\_annotation\_value}(\arg{annotation,key,value,stat})
\item\routine{get\_annotation\_value}(\arg{annotation,i,value,stat})

  This routine has two interfaces. The first gets the \arg{value}
  associated to the \arg{key}, and the second gets the \arg{value}
  associated to the \arg{i}'th entry in the annotation object.

\item\routine{get\_annotation\_key}(\arg{annotation,i,key,stat})
  
  Gets the \arg{key} of the \arg{i}'th entry in the annotation object.
  
  Together with the second form of \routine{get\_annotation\_value},
  this routine can be used to scan the complete annotation object. The
  first form of \routine{get\_annotation\_value} is appropriate if the
  key(s) are known.

\end{itemize}

In all the above routines a non-zero \arg{stat} signals an error condition.

\section{Discussion: the {\sc psml} model and ecosystem}
\label{sec:discussion}

Our vision for the role of {\sc psml} in addressing the interoperability and
documentation problems is as follows:
\begin{itemize}
  \item Databases offer {\sc psml} files (with smart searching made possible
    by the clear internal structure) and most codes use them
    directly. As they have full provenance and a uuid tag built in,
    calculations can properly document the pseudopotentials used. In
    some cases, specific legacy formats can also be produced from {\sc psml}
    files.
  \item {\sc psml} files are produced directly by most pseudopotential
    generators, either \emph{de novo} or through conversion of existing 
    formats (with extra metadata added).
  \item The {\sc psml} ideas and technology are extended to include PAW datasets
    and ultra-soft pseudopotentials.
\end{itemize}

The {\sc psml} format is able to span a wide range of uses for
norm-conserving pseudos, from semilocal-only data files, up to
full-operator datasets, with flexibility (e.g.: cascading grids),
robustness (e.g.: comprehensive exchange-correlation specification)
and full provenance. It does so in an extensible manner, and we plan to
adapt it to include PAW and USPP in the future.

To appreciate the place of {\sc psml} in relation to similar work, we can
list its strengths compared to the corresponding subset of Quantum
Espresso's UPF format (which however already supports PAW datasets and ultra-soft
pseudopotentials):

\begin{itemize}
\item The {\sc psml} format is accompanied by a complete stand-alone
  processing library that eases its adoption by client codes.
\item Support for alternative forms of datasets in the same file (i.e.,
  ``scalar-relativistic'' plus ``spin-orbit'' and/or ``lj'' form).
\item Support for different grids for different radial functions.
\item API based on interpolation that can suit any client grid,
  without having to adapt the client code to the dataset grid.
\item Full provenance specification, even when various codes are
  involved in different stages of the pseudopotential generation
  (i.e., semilocal generation followed by KB transformation).
\item A more complete specification of the exchange-correlation
  functional(s) used through a {\sc libxc}-compatible scheme.
\item A complete and parseable specification of the valence
  configuration used for generation, and of the reference
    energies used for the projectors.
\end{itemize}

The above list could serve also in a comparison of {\sc psml} with the QSO
XML-based norm-conserving pseudopotential format (See Ref.~\cite{gygi}).

As a proof of concept of the above vision, we have modified two
different atomic pseudopotential generation codes to generate {\sc psml}
files, and interfaced \libpsml{} to two electronic-structure programs.

The first generator is the open-source {\sc oncvpsp} code implemented by
D. Hamann~\cite{Hamann-13} to generate optimized multiple-projector
norm-conserving pseudopotentials. The projectors are directly stored
in the {\sc psml} format together with the local potential. In
addition, a set of semi-local potentials, a by-product of the {\sc oncvpsp}
algorithm, is also included in the {\sc psml} file. The patches needed to
produce {\sc psml} output in {\sc oncvpsp} are available in 
the Launchpad code development platform~\cite{oncvpsp-patch}.
To ease the production of XML, a special
library (\textsl{wxml}, part of the \name{xmlf90} project maintained by one of the authors
(A.G.))~\cite{xmlf90} is used.

The second generator enabled for {\sc psml} output is the {\sc atom}
code, originally developed by S. Froyen, later modified by
N. Troullier and J. L. Martins, and currently maintained by one of us
(A. G.)  within the {\sc siesta} project.  {\sc atom}, freely
distributed to the academic community, generates norm-conserving
pseudopotentials in the semilocal form. We have developed a
post-processing tool (\name{psop}) which takes as input the semilocal
components and computes a smooth local potential and the KB projector
functions in the same way as it is done within the {\sc Siesta}
code. These new elements, together with a new provenance record, are
incorporated in a new {\sc psml} file, which describes a well-defined,
client-code independent and unique operator.

We have thus already two different generators of {\sc psml} files, their
specific idiosyncrasies being describable by a common standard. Our
plans are to enable {\sc psml} output in other pseudopotential-generation
codes. 

On the client side, we have incorporated the \libpsml{} library in {\sc
  Siesta} (version 4.2, soon to be released) and {\sc Abinit} (version
8.2 and higher). {\sc psml} files can then be directly read and digested by these
codes as described in Sec.~\ref{sec:libpsml}, achieving
pseudopotential interoperability between these codes, as exemplified
in Sec.~\ref{sec:results} below.

Our immediate plans include the development of a UPF-to-{\sc psml} converter
and to encourage the adoption of \libpsml{} to enable the interoperability
of {\sc Siesta} and the codes in the Quantum Espresso suite.

We are aware that a complete deployment of the {\sc psml} road-map will
require the development of interfaces to \libpsml{} in other languages,
such as C and Python, which are in progress.

\section{Interoperability example: local-orbital and PW calculations with the
 same pseudopotential}
 \label{sec:results}

We present in this section two examples of interoperability between
{\sc siesta} and {\sc abinit} using {\sc psml} files: (i) the test of the
convergence of a numerical atomic orbital basis set 
with respect to the asymptotic limit of a
converged basis of plane waves, and (ii) the equation-of-state (energy
versus volume profiles) for elemental crystals, a test that has been
proposed as a benchmark for the comparison of different
codes~\cite{Lejaeghere-16}.

 Four paradigmatic systems are chosen as a testbed: 
 a standard semiconductor (bulk Si in the diamond structure),
 an $sp$-metal (bulk Al in the fcc structure),
 a noble metal (bulk Au in the fcc structure),
 and a $3d$ ferromagnetic transition metal (bulk Fe in the bcc structure).
 
 \begin{table*}
    \caption[ ]{ Reference configuration and cutoff radii
                 of the optimized norm-conserving pseudopotentials 
                 generated with the
                 {\sc oncvpsp} code by Hamann~\cite{Hamann-13}.
                 Units in bohr.
               }
    \begin{center}
       \begin{tabular}{cccccc}
          \hline
          \hline
                                                              &
                                                              &
          Au                                                  &
          Fe                                                  \\
          Reference                                           &
                                                              &
          $5s^{2}, 5p^{6}, 5d^{10},6s^{1}$                    &
          $3s^{2}, 3p^{6}, 3d^{6}, 4s^{2}$                    \\
          \hline
          Cutoff radius                                       &
          $s$                                                 &
          1.456                                               &
          1.163                                               \\
                                                              &
          $p$                                                 &
          1.564                                               &
          1.109                                               \\
                                                              &
          $d$                                                 &
          1.564                                               &
          1.310                                               \\
                                                              &
          $f$                                                 &
          1.602                                               &
          -                                                   \\
          Exchange-correlation                 &
                                                              &
          GGA-PBE~\cite{Perdew-96}                                &  
          GGA-PBE~\cite{Perdew-96}                                \\
          Scalar relativistic?                                &
                                                              &
          yes                                                 &
          yes                                                 \\
          Core corrections (NLCC)?                 &
                                                              &
          yes                                                 &
          yes                                                 \\
          NLCC cutoff radius                          &
                                                              &
          0.783                                               &
          0.417                                               \\
          \hline
          \hline
       \end{tabular}
    \end{center}
    \label{table:pseudopotentials-oncvpsp}
 \end{table*}

 \begin{table*}
    \caption[ ]{ Reference configuration and cutoff radii
                 of the norm-conserving pseudopotentials generated with the
                 {\sc atom} code following the Troullier-Martins
                 scheme~\cite{Troullier-91}. 
                 LDA refers to the local density approximation functional
                 of Ceperley and Alder~\cite{Ceperley-80} as parametrized
                 by Perdew and Zunger~\cite{Perdew-81}. 
                 GGA refers to the generalized gradient approximation
                 functional proposed by Perdew, Burke and 
                 Ernzerhof.~\cite{Perdew-96}
                 Units in bohr.
               }
    \begin{center}
       \begin{tabular}{cccccc}
          \hline
          \hline
                                                              &
                                                              &
          Si                                                  &
          Al                                                  &
          Fe                                                  \\
          Reference                                           &
                                                              &
          $3s^{2}, 3p^{2}$                    &
          $3s^{2}, 3p^{1}$                    &
          $4s^{2}, 3d^{6}$                    \\
          \hline
          Cutoff radius                                       &
          $s$                                                 &
          1.90                                                &
          2.30                                                &
          2.25                                                \\
                                                              &
          $p$                                                 &
          1.90                                                &
          2.30                                                &
          2.75                                                \\
                                                              &
          $d$                                                 &
          1.90                                                &
          2.30                                                &
          2.00                                                \\
                                                              &
          $f$                                                 &
          1.90                                                &
          2.30                                                &
          2.00                                                \\
          Exchange-correlation                  &
                                                              &
          LDA~\cite{Ceperley-80,Perdew-81}                    &  
          LDA~\cite{Ceperley-80,Perdew-81}                    &  
          GGA~\cite{Perdew-96}                                \\
          Relativity option                                &
                                                              &
          non-relativistic                                                  &
          non-relativistic                                                  &
          scalar-relativistic                                 \\
          Core corrections (NLCC)?                 &
                                                              &
          no                                                  &
          no                                                  &
          yes                                                 \\
          NLCC cutoff radius                          &
                                                              &
          -                                                   &
          -                                                   &
          0.70                                                \\
          \hline
          \hline
       \end{tabular}
    \end{center}
    \label{table:pseudopotentials-atom}
 \end{table*}

 The reference electronic configurations, cutoff radii, core
 corrections flags,
 and other parameters required to generate the pseudopotentials
 with the {\sc oncvpsp} code (version 3.3.0) are taken from the Pseudo-Dojo 
 database~\cite{pseudo-dojo}, and summarized in
 Table~\ref{table:pseudopotentials-oncvpsp}. 
 The pseudopotentials for Au and Fe 
 include the semicore states (5$s$ and 5$p$ for Au, 
 and 3$s$ and 3$p$ for Fe) in the valence.

 For the Troullier-Martins pseudopotentials generated with the {\sc atom}
 code, we use the parameters reported in 
 Table~\ref{table:pseudopotentials-atom}.
 The rest of the technical details of the calculations, that are common to
 both {\sc siesta} and {\sc abinit} simulations, are given in  
 Table~\ref{table:technicalities}.

 \begin{table*}[h]
    \caption[ ]{Parameters used in the simulations of bulk Si, Al and  Fe
                that are kept the same in {\sc siesta}
                and {\sc abinit}. 
                FD stands for Fermi-Dirac smearing.}
    \begin{center}
       \begin{tabular}{lllll}
          \hline
          \hline
                                                              &   
          Si                                                  &  
          Al                                                  & 
          Au                                                  &
          Fe                                                  \\
          \hline
          Atomic structure                                    &   
          Diamond                                             &  
          FCC                                                 &
          FCC                                                 &  
          BCC                                                 \\
          Exchange-correlation                 &
          LDA~\cite{Ceperley-80,Perdew-81}                    &  
          LDA~\cite{Ceperley-80,Perdew-81}                    & 
          GGA~\cite{Perdew-96}                                &
          GGA~\cite{Perdew-96}                                \\
          Spin polarized?                                     &   
          No                                                  &  
          No                                                  &  
          No                                                  & 
          Yes                                                 \\
          Monkhorst-Pack mesh                                 &   
          $6  \times 6  \times 6$                             &  
          $20 \times 20 \times 20$                            &
          $20 \times 20 \times 20$                            &
          $20 \times 20 \times 20$                            \\
          Occupation option                                   &   
          --                                                  &   
          FD                                                  &
          FD                                                  &
          FD                                                  \\
          Smearing temperature                                &   
          --                                                  &   
          0.01 Ha                                             &
          0.01 Ha                                             & 
          0.01 Ha                                             \\
          \hline
          \hline
       \end{tabular}
    \end{center}
    \label{table:technicalities}
 \end{table*}

 In {\sc siesta}, the electronic density, Hartree, and exchange
 correlation potentials, as well as the corresponding matrix elements
 between the basis orbitals, were calculated on a uniform real space
 grid, controlled by an energy cutoff~\cite{Ordejon-96} which was well
 converged at values of 300 Ry for Si, 400 Ry for Al and Fe, and 600
 Ry for Au. The reader has to keep in mind that these cutoffs are not
 directly comparable to plane-wave cutoffs, and that they control
 the operation count of stages of the calculation which typically
 represent a small fraction of the total computation time.

\begin{figure} [h]
    \begin{center}
       \includegraphics[width=\columnwidth]{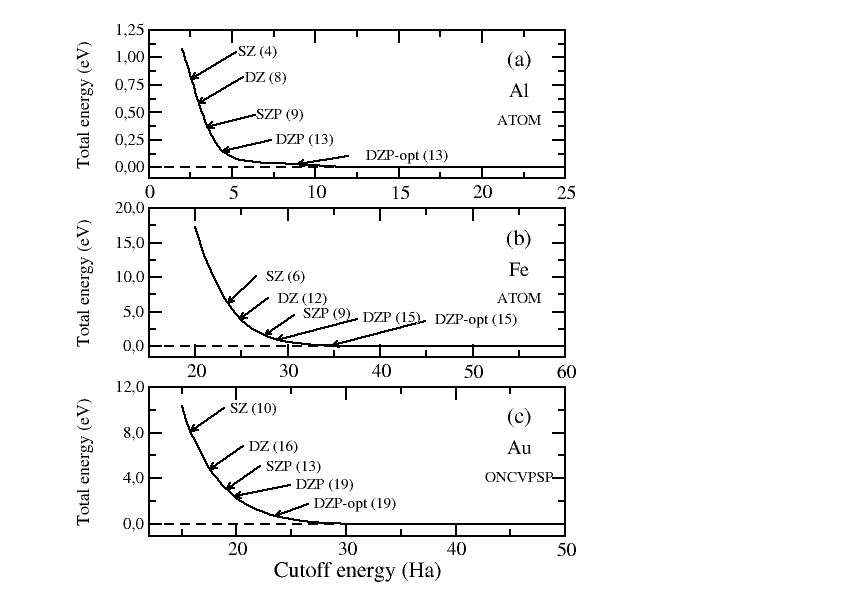}
       \caption{ Comparison of the convergence of a basis set of
         numerical atomic orbitals (NAOs) and a basis set of plane waves for
         (a) bulk Al in the fcc structure, (b) bulk Fe in the
         ferromagnetic bcc structure, and (c) bulk Au in the fcc
         structure.  The pseudopotentials for Al and Fe were generated
         within the Troullier-Martins schema as implemented in the
         {\sc atom} code, while the pseudo for Au was generated with
         the {\sc oncvpsp} code.  SZ and DZ stand, respectively, for
         single-$\zeta$ and double-$\zeta$ quality of the basis set.
         P stands for polarized.  When semicore states are included in
         the valence, the corresponding shells are always treated at
         the SZ level.  The total energy of a well converged PW
         calculation (25 Ha for Al, 60 Ha for Fe, and 50 Ha for Au)
         has been taken as the reference zero energy (dashed
         line). The number in parenthesis indicate the number of
         NAOs considered per atom.  }
       \label{fig:converener}
   \end{center}
 \end{figure}

 Figure \ref{fig:converener} shows the convergence of the total energy
 for Al, and Fe, using {\sc atom}-generated pseudopotentials, 
 and for Au using {\sc oncvpsp} pseudopotentials, as a
 function of the basis-set quality. For {\sc abinit}, the latter can 
 simply be measured by the plane-wave energy cutoff. For {\sc siesta},
 the hierarchy and the nomenclature of the basis sets of numerical atomic
 orbitals is described in Ref.~\cite{Junquera-01}.
 For a given tier within the basis hierarchy, the NAO basis sets used to 
 produce the results of Fig.~\ref{fig:converener} were generated using the default
 parameters implemented in {\sc siesta}. The only exceptions are those marked with an
 ``opt'' suffix for the {\sc atom} pseudopotentials,
 that were optimized following the recipe given in Ref.~\cite{Anglada-02}, and are 
 available on the {\sc siesta} web page~\cite{siesta-web}.
 The ``optimized'' NAO basis sets of Au and Fe with the {\sc oncvpsp} pseudopotentials 
 were generated following the automatic
 procedure implemented in {\sc siesta} with the cutoff radii for all the 
 shells defined by a unique parameter: the energy shift~\cite{Artacho-99}
 (0.005 Ry for Au and 0.002 Ry for Fe).
 The calculations are made at lattice constants of 
 3.97 \AA\ for Al, 2.87 \AA\ for Fe, and 4.16 \AA\ for Au.

 Although the convergence of NAO results is not \emph{a priori} systematic
 with respect to the size of the basis, the sequence of bases presented
 in Fig.~\ref{fig:converener} shows a uniform convergence of the
 total energy with respect the basis size.
 This is specially remarkable for fully optimized basis sets, such as the one used with
 the DZP quality for Fe and Al, where several eV can be gained.
 But significant reductions in the total energy can be obtained simply
 by tuning a reduced subset of the parameters that define the atomic orbitals.
 This is exemplified here in the case of metallic bulk Au, 
 where the total energy is lowered by almost 2 eV increasing the range of the atomic orbitals
 for the same DZP size of the basis.
 In any case, we can observe how the polarization orbitals
 are important for convergence, much more so than doubling the basis set.
 A basis of relatively modest size (DZP) is equivalent, from the
 total energy point of view, to the PW basis cutoff 
 which would be used in realistic calculations (9 Ha for Al, 35 Ha for Fe, or 23 Ha for Au).

 It is important to stress that, when using the same pseudopotential,
 the total energies are given with respect to the same reference and,
 therefore, can be \emph{directly} compared. The use of a common
 pseudopotential format thus allows a more straightforward and detailed analysis
 of convergence.
 
 In related work, a comparison of energy differences, ionic forces and
 average pressures for water monomers, dimers, two phases of ice and
 liquid water at ambient and high density have been presented in
 Ref.~\cite{Corsetti-13}. They were obtained with {\sc siesta} and
 {\sc abinit} using the same pseudopotentials (with an early version
 of the {\sc psml} framework).  Highest order bases are shown to give
 accuracies comparable to a plane-wave kinetic energy cutoff of around
 1000 eV.
 
 \begin{figure} [h]
    \begin{center}
       \includegraphics[width=\columnwidth]{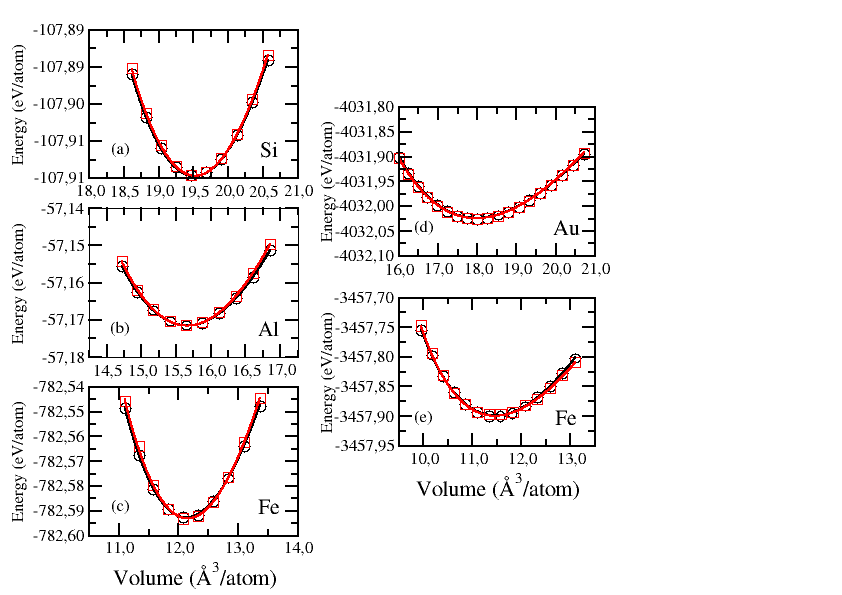}
       \caption{ (Color online) Equation of state (energy versus volume) for
                 crystalline solids using the same pseudopotential operator but 
                 different basis sets 
                 (plane waves as implemented in {\sc abinit} (black solid lines and circles),
                 and numerical atomic orbitals (DZP) as 
                 implemented in {\sc siesta} (red solid lines and 
                 squares)). 
                 Left column: Troullier-Martins norm-conserving pseudopotentials obtained
                 from the {\sc atom} code.
                 Right columne: optimized norm-conserving pseudopotential computed from
                 the {\sc oncvpsp} code.
                 The cutoff energies used in the PW calculations were
                 those corresponding to the optimized NAO basis set of DZP quality, according to 
                 Fig.~\ref{fig:converener}.
                 For {\sc atom} pseudopotentials: 13 Ha for Si, 9 Ha for Al and
                 35 Ha for Fe.
                 For {\sc oncvpsp} pseudopotentials: 51 Ha for Fe, and 23.5 Ha for Au.
                 Total energies per atom are shown to highlight the 
                 reproducibility of the results in absolute terms. 
                 }
       \label{fig:cohesioncurve}
   \end{center}
 \end{figure}
 
 Figure \ref{fig:cohesioncurve} shows the comparison of the equations
 of state for bulk Si, Al, Fe and Au, computed with a basis set of
 double-zeta polarized NAOs and for a PW basis set of comparable quality
 (see Fig.~\ref{fig:converener}). Both the position of the minimum
 energy and the curvature of the energy as a function of volume are very
 similar, indicating that for the same quality of the basis we can
 obtain essentially the same structural information. 

 The distance between the two curves, 
 for the volume range plotted in Fig.~\ref{fig:cohesioncurve}, can be 
 further quantified using the delta-factor\cite{Lejaeghere-14}. Taking the plane wave
 equation of state as a reference, we quantify the delta values
 that are reported in Table~\ref{table:structuralparam}.  In every case,
 the delta factor is smaller than 1 meV/atom, demonstrating
 the excellent agreement obtained between the two codes, and
 highlighting the level of interoperability achievable.

 \begin{table}[h]
    \caption[ ]{ Lattice constant ($a$) and bulk modulus ($B$) for 
                 bulk Si, Al, Fe and Au obtained after fitting the equation
                 of state, Fig.~\ref{fig:cohesioncurve}, to a 
                 Birch-Murnaghan equation.
                 For Fe we also show the magnetic moment ($M$) 
                 at the minimum energy structure. 
                 $\Delta$ refers to the delta 
                 factor~\cite{Lejaeghere-14,Lejaeghere-16},
                 taking the plane wave results as reference.
                 Experimental numbers from Ref.~\cite{Kittel}.}
    \begin{center}
       \begin{tabular}{llll}
          \hline
          \hline
                                  &   
          PW                      &  
          NAO                     & 
          Expt.                   \\
          \hline
          \multicolumn{4}{c}{Si}  \\
          \hline
          $a$ (\AA\ )             &          
          5.384                   &
          5.385                   &
          5.430                   \\
          $B$ (GPa)               &          
          95.7                    &
         100.2                    &
          98.8                    \\
          $\Delta$ (meV/atom)     &          
          0.93                    &
                                  &
                                  \\
          \hline
          \multicolumn{4}{c}{Al}  \\
          \hline
          $a$ (\AA\ )             &          
          3.971                   &
          3.971                   &
          4.05                    \\
          $B$ (GPa)               &          
          80.9                    &
          87.4                    &
          72.2                    \\
          $\Delta$ (meV/atom)     &          
          0.51                    &
                                  &
                                  \\
          \hline
          \multicolumn{4}{c}{Fe ({\sc atom})}  \\
          \hline
          $a$ (\AA\ )             &          
          2.895                   &
          2.896                   &
          2.870                   \\
          $B$ (GPa)               &          
          137.6                   &
          150.6                   &
          168.3                   \\
          $M$ ($\mu_{\rm B}$)     &          
          2.32                    &
          2.35                    &
          2.22                    \\
          $\Delta$ (meV/atom)     &          
          0.82                    &
                                  &
                                  \\
          \hline
          \multicolumn{4}{c}{Fe ({\sc oncvpsp})}  \\
          \hline
          $a$ (\AA\ )             &          
          2.841                   &
          2.840                   &
          2.870                   \\
          $B$ (GPa)               &          
          175.9                   &
          172.4                   &
          168.3                   \\
          $M$ ($\mu_{\rm B}$)     &          
          2.19                    &
          2.18                    &
          2.22                    \\
          $\Delta$ (meV/atom)     &          
          0.24                    &
                                  &
                                  \\
          \hline
          \multicolumn{4}{c}{Au ({\sc oncvpsp})}  \\
          \hline
          $a$ (\AA\ )             &          
          4.158                   &
          4.158                   &
          4.08                    \\
          $B$ (GPa)               &          
          139.3                   &
          139.4                   &
          173                     \\
          $\Delta$ (meV/atom)     &          
          0.14                    &
                                  &
                                  \\
          \hline
          \hline
       \end{tabular}
       \label{table:structuralparam}
    \end{center}
 \end{table}

\section{Conclusions}
We have presented the {\sc psml} norm-conserving pseudopotential file
format and the associated open source \libpsml{} library for parsing
and data handling. {\sc psml} is based on XML and implements
provenance and flexibility in a widely applicable and extensible
format. We demonstrate its potential for enabling 
interoperability among electronic-structure codes by comparing results from a
plane wave ({\sc abinit}) and an atomic orbital code ({\sc siesta}),
using the same input. We find a systematic convergence in absolute
values of energies, and a delta factor of less than 1 meV.

\section{Acknowledgements}
We thank Xavier Gonze in particular for backing this project and
providing useful comments. Many constructive discussions are
acknowledged with Don Hamann, Matteo Giantomassi,
Michiel Van Setten, Paolo Giannozzi, Gian-Marco Rignanese, and
Fran\c{c}ois Gygi.
This work was supported by CECAM through the Electronic Structure
Library (ESL) initiative and the ETSF through the libpspio project.
MJV acknowledges support from ULg and CfWB through ARC projects AIMED
and TheMoTherm (GA 15/19-09 and 10/15-03) and a FNRS PDR project (GA
T.1077.15-1/7).  
A.G. was funded by EU H2020 grant 676598 (``MaX: Materials at the eXascale''
CoE), Spain's MINECO
(grants FIS2012-37549-C05-05 and FIS2015-64886-C5-4-P, and the ``Severo Ochoa''
Program grant SEV-2015-0496), and GenCat (2014 SGR 301).
JJ and YP acknowledge support from Spain's MINECO (grants RTC-2016-5681-7 and FIS2015-64886-C5-2-P).


\end{document}